# Two fluids higher dimensional FRW cosmological model rejuvenating the cosmological tests of parametrization of Hubble Parameter in Lyra geometry

Syed Sabanam[1] and Kangujam Priyokumar singh[2]

[1,2] Department of Mathematics, Manipur University, Canchipur-795003, Imphal

Abstract

This paper has studied five-dimensional FRW cosmological models for k$= -1, 0, 1$ in the presence of two perfect fluids: ordinary baryonic fluid and a bizarre creating dark energy within the framework of Lyra's manifold (Lyra 1951). We have obtained the exact solution of the equation assuming the parametrization of the Hubble parameter H (R) = μ (1+$R^{-\vartheta}$), μ>0 and are $\vartheta$ >0 constants and the relation between metric coefficients $B = A^n$, where $n \neq 0$ is a constant. They lead to the shift from the past decelerating universe to the current accelerating universe in a time-dependent deceleration parameter. We have also constrained our proposed model parameters with 46 observational Hubble datasets. Numerous cosmological parameters have been examined about the universe's history. As such, our model can be considered a realistic one.

Keywords: Lyra geometry, Cosmological Test, FRW universe, Age of Universe, Observational data.

## 1. Introduction

While Hubble's contribute on advance empirical cosmology, Einstein's General Theory of Relativity (GR) established the groundwork for theoretical cosmology. Hubble has pointed out that galaxies are moving away from each other i.e. the universe is expanding. The present observational datas indicate that the expansion of the universe is accelerating. The universe's accelerated expansion could not be explained in the background of general relativity. To deal with those problems, several alternative theories of gravity have been studied in recent years. The scalar-tensor theories are considered the simplest and best-understood modified theory of gravitation.

Researchers from many fields have recently looked into some of the alternative theories of gravity. Among them, the most significant theories are the Weyl theory [1], Lyra geometry [2], Brans-Dicke theory [3], f(R) theory [4], and f(R, T) gravity [5] where $R$ and T are Ricci scalar and trace of the stress-energy tensor of the matter respectively.

In 1951, Lyra embarked on a transformative journey, introducing a gauge function that circumvented the nonintegrability hurdle. This pivotal modification of Riemannian geometry, known as Lyra's geometry, marked a new era of geometric exploration. Subsequently, Sen [6] built upon this foundation, putting forth a scalar-tensor theory of gravity that paralleled

Einstein's field equations, now enriched by the elegance of Lyra's geometry. Notably, Halford's work [8] shed light on the connection between the constant displacement vector field in Lyra geometry and the cosmological constant of General Relativity. His, 1972, study further revealed that the scalar-tensor treatment within Lyra's geometry produced effects that aligned, within observational constraints, with those predicted by Einstein's theory.

At present, our universe is undergoing an accelerated expansion proposed by Perlmutter et al.[8], Riess et al.[11] and Solheim [12]. WMAP and SDSS collaborations explain this fact proposed by Spergel et al. [13,14].

In the context of the Kaluza-Klein theories [15-18] the study of higher dimensional,cosmological models has obtained much importance. Many authors [19-28] formulated higher dimensional cosmological models in the general theory of relativity.

Beesham [29] studied a four-dimensional FRW cosmological model in Lyra geometry. Mohanty, Samanta Mahanta [30] constructed Kaluza-Klein FRW cosmological models in the Lyra manifold.

A lot of researchers investigated the evolutions of cosmological models filled with perfect fluids as well as viscous fluids. Recently, many researchers have been interested in investigating the two-fluid cosmological models in general relativity. The purpose of this current research is to develop a cosmological model that represents our universe as spatially homogeneous and isotropic filled with two types of fluid namely one is ordinary baryonic perfect fluid and the other one is mysterious and bizarre dark energy perfect fluid with negative pressure which is developed by a repulsive field which gives acceleration in the universe. Many researchers such as Singh and Chaubey [31] in anisotropic Bianchi-type space-times, Reddy and Kumar [32] in a scalar-tensor theory of gravitation, Naidu et al.[33] for Bianchi type-V, Amirhashchi et al.[34,35]and Saha et al.[36] for dark energy models in FRW universe have studied two fluid scenarios. Mishra et al.[37] formulated an accelerating dark energy cosmological model in two fluids with a hybrid scale factor. Goswami [38] proposed the FRW Cosmological Model in the present Perspective. Pradhan et al. [39] studied an FLRW interacting dark energy model of the universe. Manihar and Priyokumar [40] investigated the Cosmological Model Universe Consisting of Two Forms Of Dark Energy.

In this paper, from the above discussion, we have studied five-dimensional FRW cosmological models in the framework of the Lyra manifold with parametrization of the Hubble parameter and the relation between metric coefficients. In doing so we consider interacting cases only . Ade [41] pointed out that the model is shown to satisfy present observational constraints such as Planck's latest observational results. Some cosmological parameters related to the history of the universe have been studied.

Our work is structured in the following way:

Section 2 outlines the basic field equations. Section 3 provides the solutions of the field equations, while section 4 discusses the analysis of the Hubble parameter and the transition phase of the deceleration parameter. Section 5 is dedicated to an examination of the consequences. Section 6 estimates the universe's age and section 6 demonstrates the Physical properties of interacting two-fluid models estimates the universe's age. Finally, in the last section, we present our conclusion.

## 2. Basic Field Equations

Einstein's field equations based on Lyra's manifold in normal gauge , [6,7] is

$$R_{ij} - \frac{1}{2}g_{ij}R + \frac{3}{2}\emptyset_i\emptyset_j - \frac{3}{4}g_{ij}\emptyset_k\emptyset^k = -\frac{8\pi G}{c^4}T_{ij} \qquad (1)$$

where $\emptyset_i$ denotes the Lyra displacement and other symbols have their usual meaning in the Riemannian geometry and

$$\emptyset_i(0,0,0,0,\beta(t)) \qquad (2)$$

Here we consider FRW five-dimensional space-time metric (in units $c = 1$) in the form

$$ds^2 = dt^2 - A^2(t)\left[\frac{dr^2}{1-kr^2} + r^2(d\theta^2 + \sin^2\theta\, d\emptyset^2)\right] - B^2(t) \qquad (3)$$

where t is cosmic time, A(t) is the scale factor, B is the function of t, and k=-1,0,1 for negative curvature, vanishing, space of positive representing open, flat, and closed models of the universe. The fifth coordinate $\varphi$ is also assumed to be a space-like coordinate.

The energy-momentum tensor for the two perfect fluid [38] has the form

$$T_{ij} = T_{ij}(m) + T_{ij}(de) \qquad (4)$$

Where $T_{ij}(m)$ is the energy-momentum tensor for the ordinary baryonic fluid with pressure $p_m$ and energy density $\rho_m$ and $T_{ij}(de)$ is the energy-momentum tensor for a bizarre type of fluid creating dark energy with pressure $p_{de}$ and energy density $\rho_{de}$ which are respectively given by

$$T_{ij}(m) = (\rho_m + p_m)u_iu_j - p_m g_{ij} \qquad (5)$$

$$T_{ij}(de) = (\rho_{de} + p_{de})u_iu_j - p_{de} g_{ij} \qquad (6)$$

The five velocity vectors are $u^i = (0,0,0,0,1)$ with $u^i u_i = 1$.

The field equation (1) together with (4), and (5) for the metric (6) can be written as

$$\frac{\dot{A}^2}{A^2} + 2\frac{\ddot{A}}{A} + 2\frac{\dot{A}\dot{B}}{AB} + \frac{\ddot{B}}{B} - \frac{3}{4}\beta^2 = 8\pi G(p_m + p_{de} + p_k) \tag{7}$$

$$\frac{3\ddot{A}}{A} + \frac{3\dot{A}^2}{A^2} + \frac{2k}{A^2} - \frac{3}{4}\beta^2 = 8\pi G(p_m + p_{de} + p_k) \tag{8}$$

$$\frac{3\dot{A}^2}{A^2} + \frac{3\dot{A}\dot{B}}{AB} + \frac{3}{4}\beta^2 = -8\pi G(\rho_m + \rho_{de} + \rho_k) \tag{9}$$

Where an overdot means differentiation with respect to t. For this we assume that the density and pressure for curvature energy are as follows:

$$\rho_k = \frac{3k}{8\pi G A^2} \text{ and } p_k = \frac{-k}{8\pi G A^2}$$

From equations (7), (8), and (9), we have

$$\frac{\dot{A}^2}{A^2} + 2\frac{\ddot{A}}{A} + 2\frac{\dot{A}\dot{B}}{AB} + \frac{\ddot{B}}{B} - \frac{3}{4}\beta^2 = 8\pi G p \tag{10}$$

$$\frac{3\ddot{A}}{A} + \frac{3\dot{A}^2}{A^2} + \frac{2k}{A^2} - \frac{3}{4}\beta^2 = 8\pi G p \tag{11}$$

$$\frac{3\dot{A}^2}{A^2} + \frac{3\dot{A}\dot{B}}{AB} + \frac{3}{4}\beta^2 = -8\pi G \rho \tag{12}$$

Where $p = p_m + p_{de} + p_k$, $\rho = \rho_m + \rho_{de} + \rho_k$, $p$ is total pressure, $\rho$ is total energy density of the universe.

The energy conservation law $T^{ij}_{;j} = 0$

$$\dot{\rho} + \left(\frac{3\dot{A}}{A} + \frac{\dot{B}}{B}\right)(\rho + p) = 0 \tag{13}$$

$$\dot{\rho}_k + \left(\frac{3\dot{A}}{A} + \frac{\dot{B}}{B}\right)(p_k + \rho_k) = 0 \tag{14}$$

or

$$\dot{\rho}_k + 3\left(\frac{\dot{A}}{A} + \frac{\dot{B}}{3B}\right)(p_k + \rho_k) = 0 \tag{15}$$

$$\frac{d}{dt}(\rho_m + \rho_{de}) + 3\left(\frac{\dot{A}}{A} + \frac{\dot{B}}{3B}\right)(p_m + p_{de} + \rho_m + \rho_{de}) = 0 \tag{16}$$

And

$$\left(R^j_i - \frac{1}{2}Rg^j_i\right)_{;j} + \frac{3}{2}(\emptyset_i \emptyset^j)_{;j} - \frac{3}{4}(g^j_i \emptyset_k \emptyset^k)_{;j} = 0 \tag{17}$$

Gives

$$\frac{3}{2}\beta\dot{\beta} + \frac{3}{2}\beta^2\left(\frac{3\dot{A}}{A} + \frac{\dot{B}}{B}\right) = 0 \qquad (18)$$

The average scale factor (R(t)) and spatial volume (v) of the FRW spacetime are defined by

$$v = R^4 = A^3 B \qquad (19)$$

The scalar expansion ($\theta$) is given by

$$\theta = \frac{3\dot{A}}{A} + \frac{\dot{B}}{B} \qquad (20)$$

## 3. Solution of the Field Equation

Since there are 3 (three) highly nonlinear independent Eqns. (7-9) involving 5 (five) unknown variables (A, B, $\beta$, $\rho$, and p), therefore to get determinate solutions of the above system of Eqns. we need 2 (two) extra Eqns. comprising them. To obtain the 2 (two) extra Eqns., we assume the following two physically plausible conditions.

(1) First we assume the relation between metric coefficients as considered by Singh et al.[31]

$$B = A^n \qquad (21)$$

where $n \neq 0$ is a constant.

(2) Banerjee and Das [43] proposed the parametrization of the Hubble parameter, while Singh[42] and Pacif et al.[44] assume a relation is given by

$$H(R) = \mu\,(1+R^{-\vartheta}) \qquad (22)$$

where $\mu > 0$ and $\vartheta > 1$ are constants, R(t) is an average scale factor.

Equation (22) readily shows the explicit form of R as

$$R(t) = \left(k_1 e^{\mu\vartheta t} - 1\right)^{\frac{1}{\vartheta}} \qquad (23)$$

Where $k_1$ is the constant of integration.

Using the relation between the redshift and the average scale factor of the Universe $R(t) = (1+z)^{-1}$, we can define the relation between the cosmic time and redshift as

$$t = \frac{1}{\mu\vartheta}\log\left[\frac{1+(1+z)^{-\vartheta}}{k_1}\right] \qquad (24)$$

The cosmic time t expressed in redshift z is represented by the equation above. This might be considered the current era of the universe, which is $z \to \infty$, then $t \to 0$, and as $z \to 0$ then $t \to t_0$.

We have found that

$$8\pi G p(z) = -\frac{3}{4}\beta_0^2(1+z)^8 + \frac{4\mu^2\left(1+(1+z)^{-\vartheta}\right)\left(-\vartheta(2+n)(3+n)\right)+4\left(3+n(2+n)\right)\left(1+(1+z)^{-\vartheta}\right)}{(3+n)^2\left(-1+k_1\left(\frac{1+(1+z)^{-\vartheta}}{k_1}\right)^{\frac{1}{\mu\vartheta}}\right)^2} \quad (25)$$

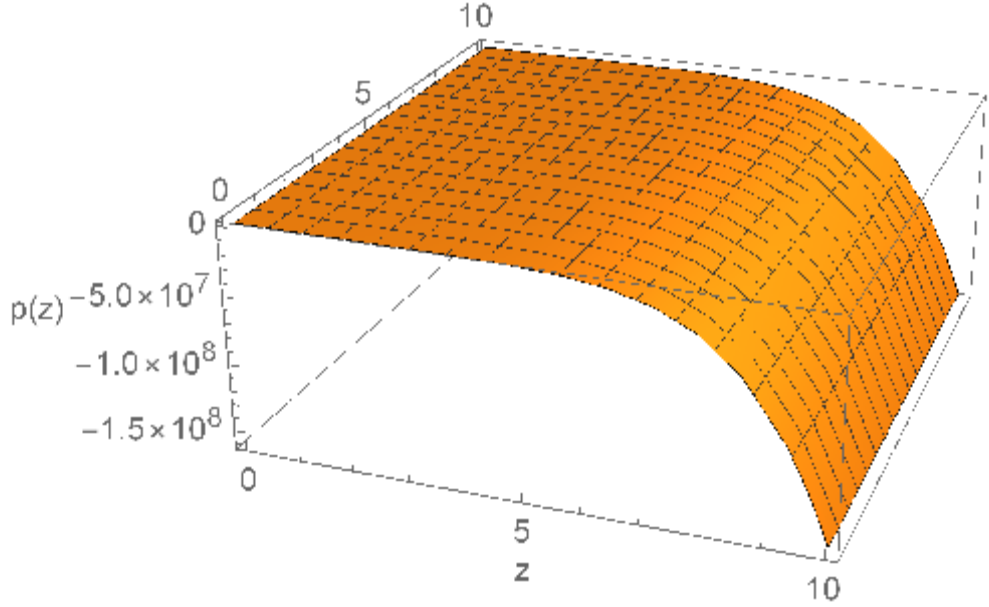

Fig.1: Plot of pressure (p) versus redshift (z) for , $\beta_0 = 1$, $\mu = 33.75$, $\vartheta = 1.209$ and $8\pi G = 1$.

Fig.(1) represents that the pressure (p) versus redshift (z). It shows that the pressure decreases continuously with increasing redshift, indicating that in the early universe at higher redshifts, the pressure was higher. At present $z = 0$, the pressure is negative. This is consistent with the present accelerated universe expansion requiring negative pressure dark energy. The steep negative slope indicates that pressure dropped rapidly in the later stages of cosmic evolution.

$$8\pi G \rho(z) = -\frac{3}{4}\beta_0^2(1+z)^8 + \frac{48\mu^2(1+n)(1+z)^{2\vartheta}\left(1+(1+z)^{-\vartheta}\right)^2}{(3+n)^2} \quad (26)$$

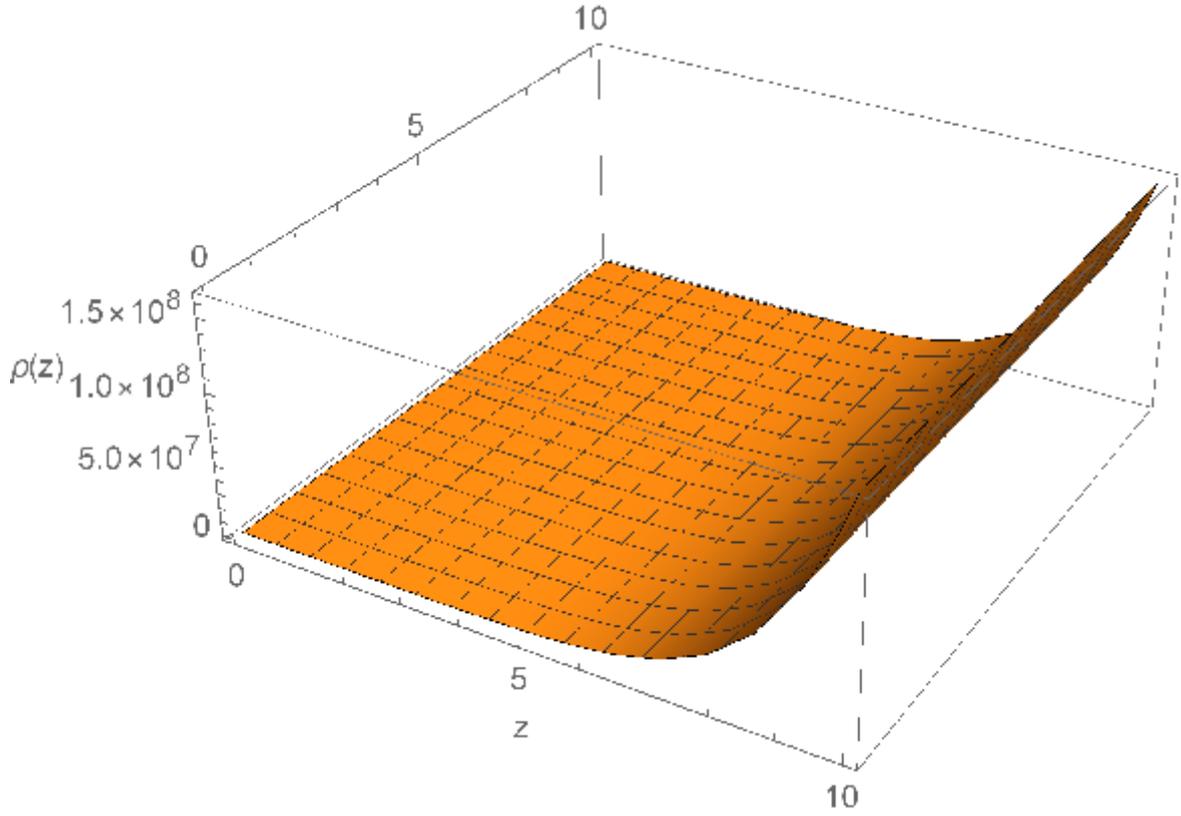

Fig.2: Plot of energy density ($\rho$) versus redshift ($z$) for n=1, $\beta_0 = 1$, $\mu = 33.75$, $\vartheta = 1.209$ and $8\pi G = 1$.

Fig. (2) demonstrates that the energy density ($\rho$) of the universe increases with redshift ($z$) and still seems to as the universe expands.

$$8\pi G(p_m + p_{de})$$
$$= -\frac{3}{4}\beta_0^2(1+z)^8 + k(1+z)^{\frac{4}{3+n}}$$
$$+ \frac{4\mu^2\big(1+(1+z)^{-\vartheta}\big)\big(-\vartheta(2+n)(3+n)\big) + 4\big(3+n(2+n)\big)\big(1+(1+z)^{-\vartheta}\big)}{(3+n)^2\left(-1 + k_1\left(\frac{1+(1+z)^{-\vartheta}}{k_1}\right)^{\frac{1}{\mu\vartheta}}\right)^2} \quad (27)$$

$$8\pi G(\rho_m + \rho_{de})$$
$$= \frac{3}{4}\beta_0^2(1+z)^8 - 3k(1+z)^{\frac{4}{3+n}}$$
$$+ \frac{48\mu^2(1+n)(1+z)^{2\vartheta}(1+(1+z)^{-\vartheta})^2}{(3+n)^2} \quad (28)$$

## 4. Analysis of Hubble Parameter

The mean Hubble parameter in terms of redshift is given by

$$H(z) = \mu(1 + (1+z)^{\vartheta}) \qquad (29)$$

$$H(z=0) = H_0 = 67.5 \text{ km } s^{-1}Mpc^{-1} \qquad (30)$$

Lastly, the Hubble rate function may be written as

$$H(z) = \frac{H_0}{2}(1 + (1+z)^{\vartheta}) \qquad (31)$$

Where $H_0 = 100h_0$ is the Hubble rate at the present time. In our model, when $\mu = 33.75$, $\vartheta = 1.209$ then $H_0 = 67.5 \text{ km } s^{-1}Mpc^{-1}$ which aligns closely with Planck results (2018) gives the present value of Hubble's constant in the range $H_0 = 67.4 \pm 0.5 \text{ km } s^{-1}Mpc^{-1}$.

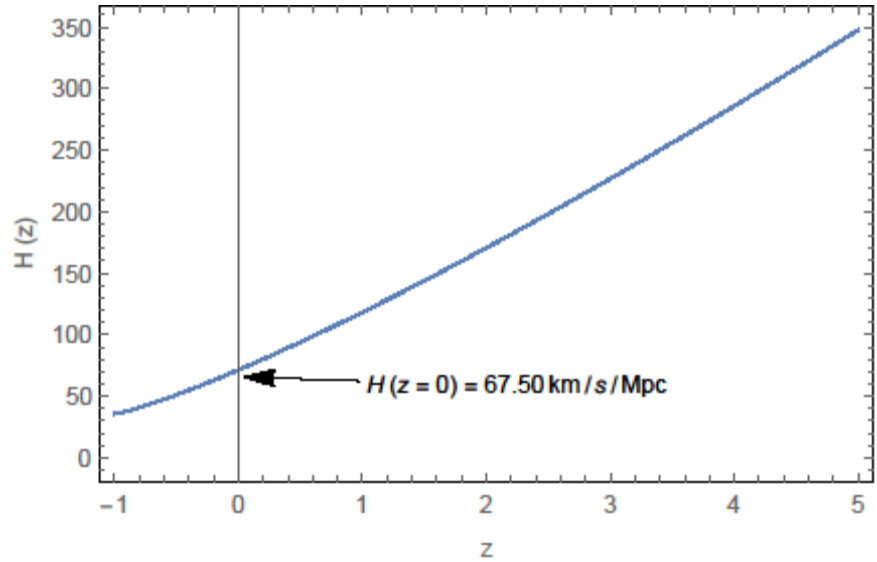

Fig.3: Plot of Hubble parameter (H) versus redshift (z) for $\mu = 33.75$, and $\vartheta = 1.209$.

Fig. (3) plots the Hubble parameter (H) against redshift (z). It shows that H increases with z.

## 4.1. Hubble Constant at Present

We consider 46 Hubble constant datasets of H(z) across redshift z with errors in H(z) that are observed in Pradhan et al. [46]. These observed values are acquired by various researchers independently. We compare these values with the theoretical results obtained by Eq. (31).

For this, we employed $x^2$-test formula is given below

$$x^2 = \sum_{i=1}^{N} \frac{[H^{th}(z_i) - H^{obs}(z_i)]^2}{\sigma^2_{H(z_i)}} \qquad (32)$$

Where N, $H^{th}(z_i)$, $H^{obs}(z_i)$ represented the total amount of data, the Hubble parameter with the model parameters and observed Hubble parameter values, respectively. $\sigma^2_{H(z_i)}$ is the standard deviation obtained from observations. We constrain the model parameters by performing a $x^2$ minimization fit to 46 observed Hubble parameter data points over redshift $0 \leq z \leq 2.36$. The best-fit yields a present Hubble constant $H_0 = 67.4 \pm 0.5$ km $s^{-1} Mpc^{-1}$ and $x^2{}_{minimum} = 0.39$ approximate.

## 4.2. Transition Phase from Deceleration to Acceleration

$$q(z) = -1 + \frac{\vartheta}{1 + (1+z)^\vartheta} \qquad (33)$$

One of the cosmological parameters that is significant in explaining the state of the expansion of our universe is the deceleration parameter q. When the value of the deceleration parameter is strictly less than zero, it shows the accelerating behavior of the universe and when it is zero or positive, the universe decelerates. Recently, Camarena and Marra [47] formulated its value $q_0 = -0.55$, whereas Capozziello et al. [48] predict the value as $q_0 = -0.644 \pm 0.223$ and $q_0 = -0.6401 \pm 0.187$. Since $q_0$ lies in the range $-1 \leq q_0 < 0$, the acceleration model universe undergoes exponential expansion, in understanding the present cosmology. In this paper, we have constrained when $\vartheta = 1.209$ the present value of deceleration parameter $q_0 = -0.40$
.

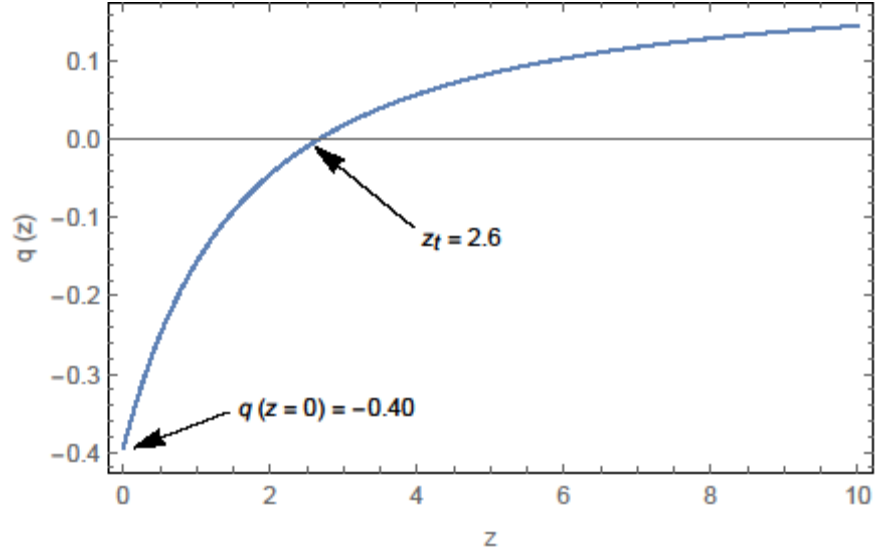

Fig.4: Plot of deceleration parameter (q) versus redshift (z) for $\vartheta = 1.209$.

Fig. (4) indicates its geometrical behavior over redshift (z). We have seen that q represents shifting from positive to negative values over its evolution with redshift at $z_t = 2.6$. This shows that our model universe is in an accelerating phase.

## 5. Some Consequences in Cosmology

### 5.1. Lookback time-redshift

The lookback time is defined as the difference between the present age of the universe $t_0$ and the age of the universe when a particular light ray at redshift z was emitted. It is defined by

$$t - t_0 = \int_R^{R_0} \frac{dR}{\dot{R}} \qquad (34)$$

Where $t_0$ is the universe's present age.

The scale factor R in terms of redshift parameter z is shown below

$$\frac{R}{R_0} = \frac{1}{1+z} \qquad (35)$$

Where $R_0$ is the present day scale factor of the universe and z denotes the redshift of light.

Using Eq. (35) in Eq.(34), we have

$$H_0(t - t_0) = \frac{2}{\vartheta}\left(\log\left[\frac{1+(1+z)^{-\vartheta}}{2}\right]\right) \qquad (36)$$

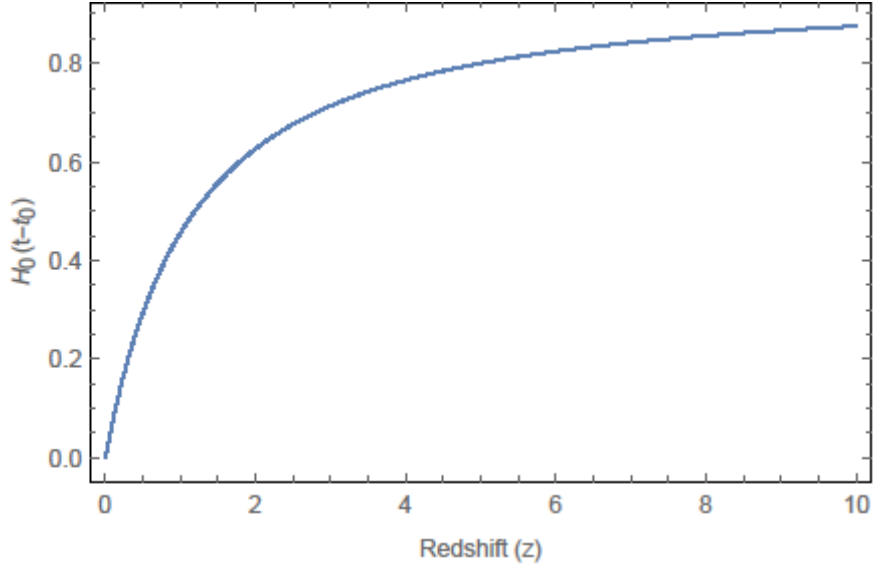

Fig.(5): Plot of lookback time $H_0(t - t_0)$ versus redshift (z) for $\mu = 33.75$, $\vartheta - 1.209$ and $k_1 = -1.9$.

Fig. (5) shows that as we go to higher redshifts, the lookback time increases continuously. Thus, light from older cosmic epochs with higher z values has traveled over longer durations before reaching us. When $z \to \infty$, $H_0(t - t_0) \to 0$ which shows expanded the universe. The lookback time initiates at the Big Bang.

5.2. Proper Distance

The proper distance d(z) is defined as the distance between a cosmic source emitting light at any instant $t = t_1$ located at $r = r_1$ with redshift $z$ and the observer receiving the light from the source emitted at $r = 0$ and $t = t_0$. It is defined by

$$d(z) = r_1 R_0 \qquad (37)$$

Where $r_1 = \int_{t_1}^{t_0} \frac{dt}{R}$

Using Eq.(35) in Eq.(37), we have

$$d(z) = \log\left[\frac{2}{H_0 \vartheta} \log\left(\frac{1+(1+z)^{-\vartheta}}{2}\right)\right] \qquad (38)$$

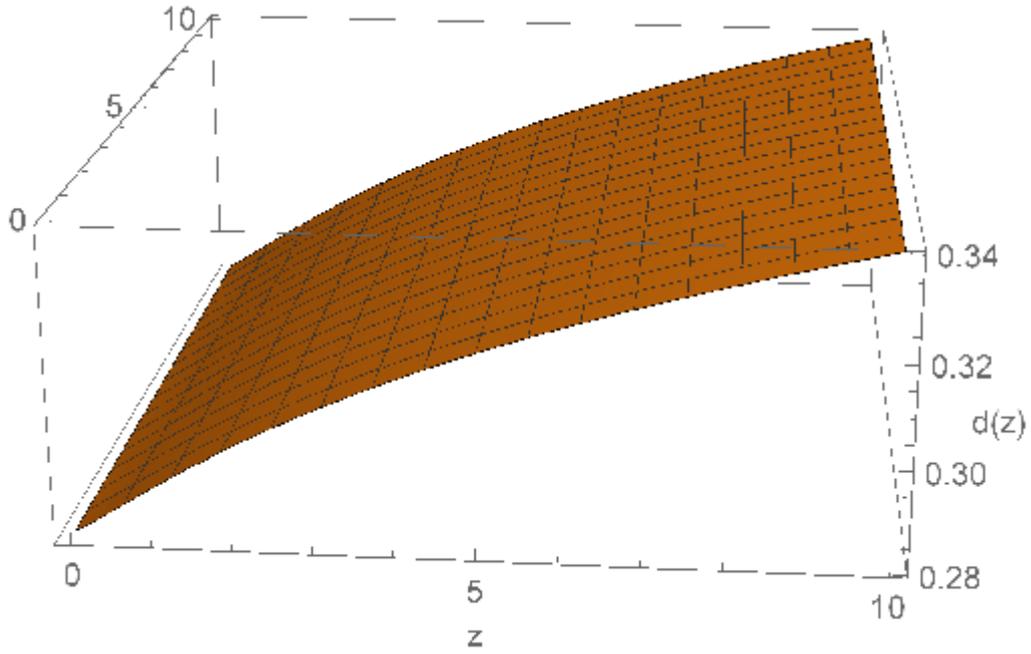

Fig. (6): Plot of proper distance d(z) versus redshift (z) for $\mu = 33.75$, $\vartheta - 1.209$ and $k_1 = -1.9$.

Fig. (6) shows the variation of proper distance (d) with redshift (z). We observe that the proper distance increases with increasing z. Thus, light sources at higher redshifts are further away in terms of properly defined cosmological distances. When $z \to \infty$, $d(z) \to 0$ is expanding. It begins with the Big Bang.

5.3. Luminosity Distance

The luminosity distance $d_L$ of light source is defined as

$$d_L = R_0 r_1 (1+z) = d(z)(1+z) \qquad (39)$$

By the use of Eq.(38) in Eq.(39), we have

$$d_L = \log\left[\frac{2}{H_0 \vartheta} \log\left(\frac{1+(1+z)^{-\vartheta}}{2}\right)\right](1+z) \qquad (40)$$

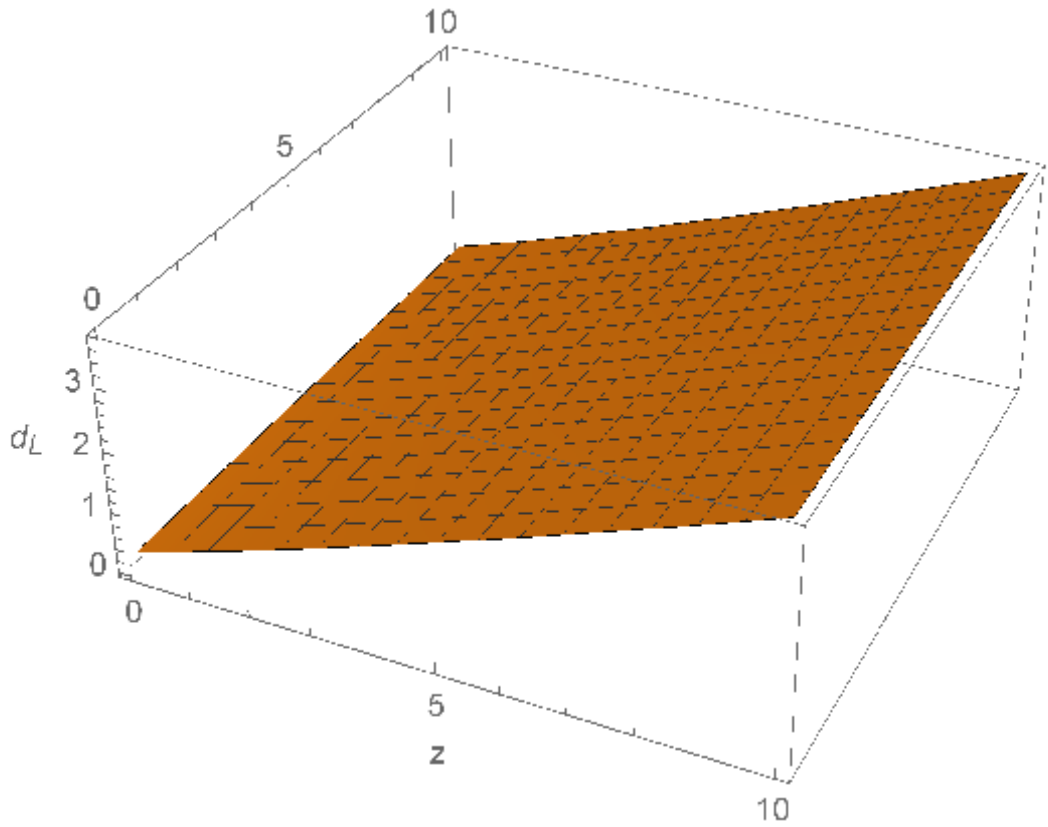

Fig.(7) Plot of luminosity distance $d_L$ versus redshift (z) for $\mu = 33.75$, $\vartheta - 1.209$ and $k_1 = -1.9$.

Fig. (7) represents that the luminosity distance $d_L$ increases continuously with redshift. Sources with higher redshift appear more luminous since the greater spatial expansion causes light dilution over larger distances. Thus, more distant objects at higher redshifts have higher luminosity distances in an expanding FRW universe. Our proposed model is a good understanding of the expansion of the universe.

5.4. Distance Modulus

The distance modulus $\mu(z)$ is defined as

$$\mu(z) = 5 \log d_L + 25 \tag{41}$$

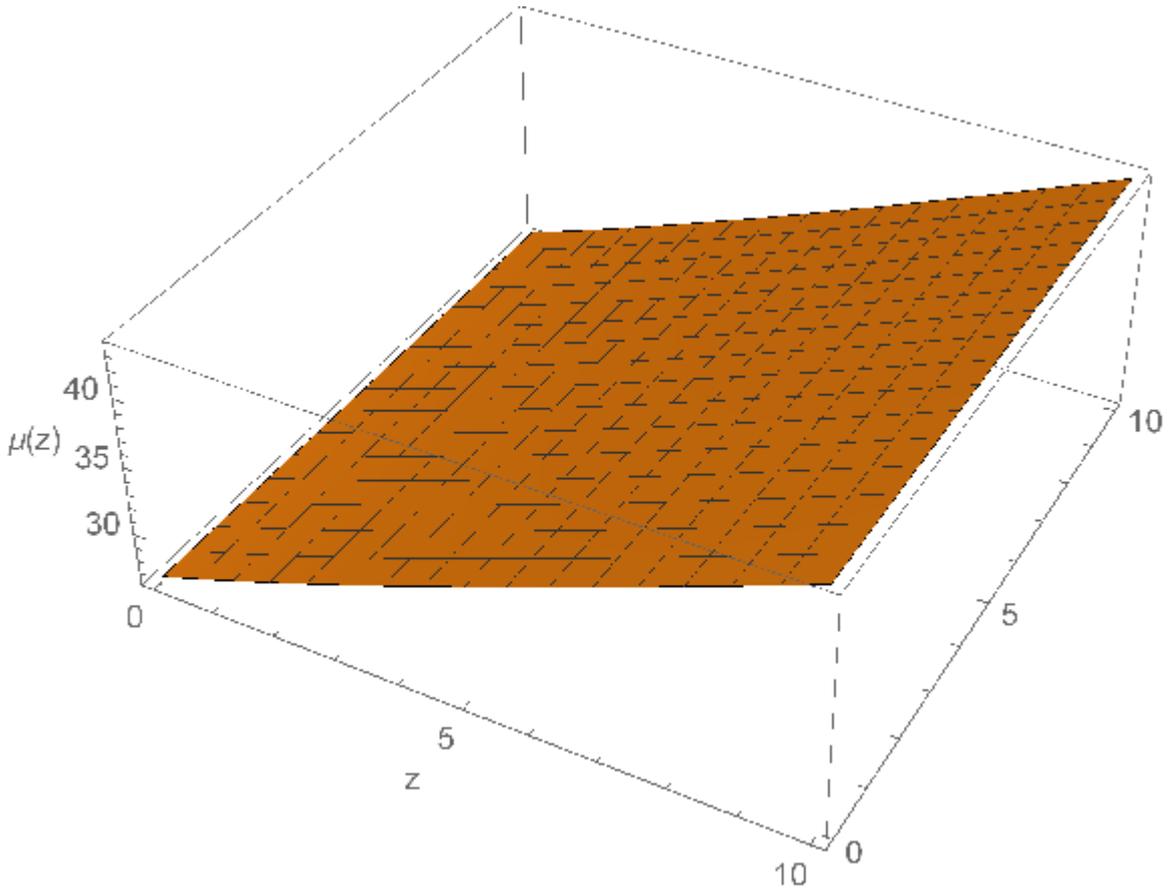

Fig.(8): Plot of distance modulus $\mu(z)$ versus redshift (z) for $\mu = 33.75, \vartheta - 1.209$ and $k_1 = -1.9$.

In Fig. (8), the distance modulus $\mu$ is plotted against redshift. Distance modulus quantifies light dimming from distant sources due to spatial expansion. We see $\mu$ increasing with z, indicating that sources at higher z appear more dimmed than those at low z, as expected in an expanding universe model causing light to travel larger distances.

## 6. Age of the Present Universe

The present age of the Universe in the proposed model is given by

$$t_0 - t = \int_0^z \frac{dz}{(1+z)H(z)} \qquad (42)$$

Using Eq.(31) in Eq.(42), we have

$$H_0(t_0 - t) = \lim_{z \to \infty} \int_0^z \frac{2dz}{(1+z)(1+(1+z)^\vartheta)} \qquad (43)$$

We can find that as $z \to \infty, H_0(t - t_0)$ tends to constant value that represents the cosmic age of the universe, $H_0 t_0 = 0.96$. The universe's current cosmic age is calculated to be $t_0 = 13.4$

Gyrs which are extremely near to estimated values based on observations when $\vartheta = 1.209, H_0 = 67.5$ km$s^{-1}Mpc^{-1}$ and $k_1 = -1.9$.

## 7. Physical Properties of Interaction Two-Fluid Models

Now, we consider the interaction between dark and baryonic fluids. For this, we can write the equation of continuity for the dark and baryonic fluids as

$$\dot{\rho}_m + \left(\frac{3\dot{A}}{A} + \frac{\dot{B}}{B}\right)(p_m + \rho_m) = Q \qquad (44)$$

And

$$\dot{\rho}_{de} + \left(\frac{3\dot{A}}{A} + \frac{\dot{B}}{B}\right)(p_{de} + \rho_{de}) = -Q \qquad (45)$$

We assume $Q \geq 0$ since the quantity Q represents the energy transfer from dark energy to baryonic matter.

If dark energy is non-interacting, Q=0.

We assume $Q > 0$, meaning that the second law of thermodynamics is satisfied since we are interested in an energy transfer from dark energy to dark matter. Thus, multiplying the Hubble factor H by the energy density can be a first and obvious candidate.

Following Amendola et al.[49] and Gou et al.[50], we assume that

$$Q = 3H\sigma\rho_m \qquad (46)$$

Where $\sigma$ is a coupling constant and is positive.

At present, our universe is dust-filled, so we consider $p_m = 0$.

Integrating Eq.(29) with the help of Eqs.(42) and (44), we have

$$\rho_m = (\rho_m)_0 \left(-1 + k_1 e^{\mu\vartheta t}\right)^{\frac{4(3\sigma-1)}{\vartheta}} \qquad (47)$$

Where $(\rho_m)_0$ is constant of integration.

In terms of redshift

$$\rho_m = (\rho_m)_0 (1+z)^{4(1-3\sigma)} \qquad (48)$$

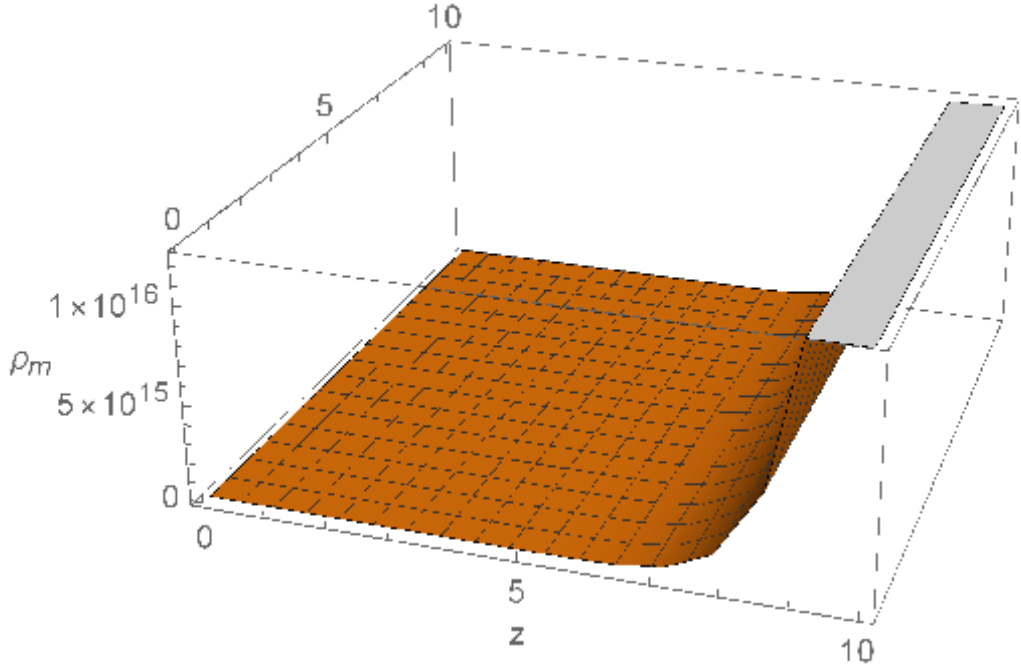

Fig.9: Plot of $\rho_m$ versus redshift (z) for, $(\rho_m)_0 = 1, \vartheta = 1.209$ and $\sigma = -1$.

Fig. (9) indicates the variation of the energy density of baryonic matter $\rho_m$ with redshift z. We observe that $\rho_m$ increases continuously with increasing z. This is expected since in the past epochs of the universe at higher redshifts, the density of ordinary baryonic matter was higher compared to present times. The matter density has decreased over cosmic expansion history as the universe evolves.

By the use of Eq.(46) in Eq.(27), we have

$$8\pi G \rho_{de} = \frac{3}{4}\beta_0^2(1+z)^8 - 3k(1+z)^{\frac{4}{3+n}} + \frac{48\mu^2(1+n)(1+z)^{2\vartheta}\left(1+(1+z)^{-\vartheta}\right)^2}{(3+n)^2}$$

$$-8\pi G (\rho_m)_0 (1+z)^{4(1-3\sigma)} \qquad (49)$$

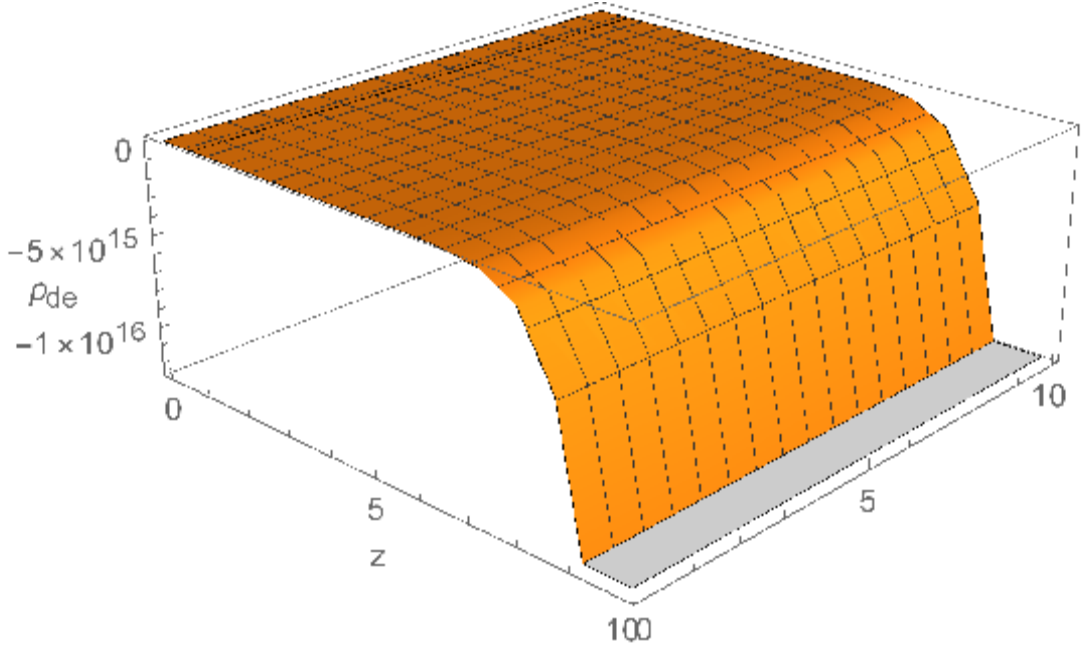

Fig.10: Plot of $\rho_{de}$ versus redshift (z) for, $(\rho_m)_0 = 1, \vartheta = 1.209, \beta_0 = 1$ and $\sigma = -1$.

From Fig. (10), we observe that the energy density of dark energy $\rho_{de}$ is negative for flat, open, and closed cases. This figure shows the energy density of dark energy $\rho_{de}$ as a function of redshift. We see that $\rho_{de}$ remains negative across the redshift range. A negative energy density for dark energy is consistent with its repulsive gravity property that drives the acceleration of the cosmic expansion. The increasingly negative values at higher redshifts indicate that in the past, the proportion of mysterious dark energy having negative pressure was higher compared to the current epoch.

$$8\pi G p_{de} = \frac{3}{4}\beta_0^2(1+z)^8 + k(1+z)^{\frac{4}{3+n}} + \frac{4\mu^2\left(1+(1+z)^{-\vartheta}\right)\left(-\vartheta(2+n)(3+n)\right)+4\left(3+n(2+n)\right)(1+(1+z)^{-\vartheta})}{(3+n)^2\left(-1+k_1\left(\frac{1+(1+z)^{-\vartheta}}{k_1}\right)^{\frac{1}{\mu\vartheta}}\right)^2} \quad (50)$$

Here, the energy densities of dark energy $\Omega_{de}$ can be written as

$$\Omega_{de} = \frac{8\pi G \rho_{de}}{2H^2} \quad (51)$$

$$\Omega_{de} = \frac{\frac{3}{4}\beta_0^2(1+z)^8 - 3k(1+z)^{\frac{4}{3+n}} + \frac{48\mu^2(1+n)(1+z)^{2\vartheta}(1+(1+z)^{-\vartheta})^2}{(3+n)^2} - 8\pi G(\rho_m)_0(1+z)^{4(1-3\sigma)}}{2\left(\mu\left(1+(1+z)^{-\vartheta}\right)\right)^2} \quad (52)$$

## 8. Conclusion

In this study, we have presented a notable solution within the context of the five-dimensional world, as described by the Kaluza-Klein theory. Our investigation encompasses a universe governed by an energy-momentum tensor of two fluids (baryonic and dark energy). We have investigated the field of the Friedmann-Robertson-Walker (FRW) model, specifically focusing on its characteristics within the Lyra geometry. Given the observed present cosmic acceleration and to attain an exact solution of the cosmological field equations, we have assumed a simple parametrization of Hubble parameter H and the relation between the metric coefficient. These result in a time-dependent transition phase from decelerated to accelerated expansion. We have studied only the interaction of two fluid models.

The main features of our proposed model are as follows:

(1) The Hubble parameter H has been parametrized geometrically, as used by Singh [42], Banerjee et al.[43], and Pacif et al. [44]. This leads to a time-dependent deceleration parameter q, which can explain the universe's current acceleration $q < 0$ with a previous deceleration $q > 0$ in the past. As a notable deviation from the Standard Model, it has been noted that the universe did not begin with a Big Bang scenario but rather with a finite volume, velocity, and acceleration.
(2) The present value of Hubble parameter is measured as $H_0$=67.5 km$s^{-1}Mpc^{-1}$.
(3) Furthermore, we used the Hubble datasets containing 46 data points to determine the best-fit values for the model parameters as $x^2{}_{minimum} = 0.39$ approximate.
(4) The present value of the deceleration parameter is measured as $q_0 = -0.40$ which suggests that our model universe expansion is in an accelerating phase supported by [10, 11, 46].
(5) We have estimated the transition $z_t$ =2.6 which very closed to Pradhan et al. [39].
(6) We have obtained expressions for lookback time-redshift, proper distance, luminosity distance, and distance modulus versus redshift and discussed their significance.
(7) We analyzed the evolution of the various cosmological parameters corresponding to these best-fit values of the model parameters.
(8) The cosmic age of the present universe is approximated as $t_0 = 13.4$ Gyrs.
(9) Lastly, we conclude that the considered parametric form of the Hubble parameter in the framework of Lyra geometry plays an important role in causing the accelerated expansion of the universe.